\documentstyle[prl,psfig,twocolumn,aps]{revtex}
\begin{document}

\title{Nonlocal Boundary Dynamics of Traveling Spots in a Reaction-Diffusion 
System}
\author{L.M. Pismen \\
{\it Department of Chemical Engineering and 
Minerva Center for Nonlinear Physics of Complex Systems,\\
Technion -- Israel Institute of Technology, 32000 Haifa, Israel}} 
\date{\today}
\maketitle

\begin{abstract}
The boundary integral method is extended to derive closed integro-differential 
equations applicable to computation of the shape and propagation speed of a 
steadily moving spot and to the analysis of dynamic instabilities in the sharp 
boundary limit. Expansion of the boundary integral near the locus of traveling 
instability in a standard reaction-diffusion model proves that the bifurcation 
is supercritical whenever the spot is stable to splitting, so that propagating 
spots can be stabilized without introducing additional long-range variables. 
\end{abstract}

\pacs{82.40.Bj, 82.20.Mj, 05.65.+b}

Localized structures in non-equlibrium systems (dissipative solitons) have been 
studied both in experiments and computations in various applications, including 
chemical patterns in solutions \cite{swin} and on surfaces \cite{imb}, gas 
discharges \cite{pur99} and nonlinear optics \cite{firth}. 
The interest to dynamic solitary 
structures, in particular, in optical \cite{firth} and gas discharge systems  
\cite{purw} has been recently driven by their possible role in information 
transmission and processing. 

A variety of observed phenomena can be reproduced qualitatively with the help of 
simple reaction-diffusion models with separated scales 
\cite{ohta,keros,meron,pi94,osmur}. Extended models of this type included 
nonlocal interactions due to gas transport \cite{pi94,mikh}, Marangoni flow 
\cite{pi97} or optical feedback \cite{firth,piop}. A great advantage of scale 
separation is a possibility to construct analytically strongly nonlinear 
structures in the sharp interface limit. An alternative approach based on 
Ginzburg--Landau models supplemented by quintic and/or fourth-order differential 
(Swift-Hohenberg) terms \cite{rab} have to rely on numerics in more than one 
dimension. 

Dynamical solitary structures are most interesting from the point of view of 
both theory and potential applications. Existence of traveling spots in 
sharp-interface models is indicated by translational 
instability of a stationary spot 
\cite{mikh}. This instability is a manifestation of a general phenomenon of parity 
breaking (Ising--Bloch) bifurcation \cite{coul,hagmeron} which takes a single 
stable front into a pair of counter-propagating fronts forming the front and the 
back of a traveling pulse. Numerical simulations, however, failed to produce 
stable traveling spots in the basic activator-inhibitor model, and the tendency 
of moving spots to spread out laterally had to be suppressed either by global 
interaction in a finite region \cite{mikh} or by adding an extra inhibitor with 
specially designed properties \cite{bode}.

The dynamical problem is difficult for theoretical study, since a moving spot 
loses its circular shape, and a free-boundary problem is formidable even for 
simplest kinetic models. Numerical simulation is also problematic, due to the 
need to use fine grid to catch sharp gradients of the activator; therefore 
actual computations were carried out for moderate scale ratios.  
A large amount of numerical data, such as the inhibitor field far from the spot 
contour, is superfluous. This could be overcome if it was possible 
to reduce the PDE solution to 
local dynamics of a sharp boundary. Unfortunately, a purely local equation of 
front motion \cite{hagmeron} is applicable only when the curvature far exceeds 
the diffusion scale of the long-range variable, whereas a spot typically  
suffers splitting instability \cite{ohta} before growing so large. On the other 
hand, the nonlocal boundary integral method \cite{gope} is applicable only when 
the inhibitor dynamics is fast compared to the characteristic propagation scale 
of a front motion, i.e. under conditions when no dynamic instabilities arise and
traveling spots do not exist.

It is the aim of this Letter, to extend the nonlocal boundary integral method  
to dynamical problems, and to find out with its help conditions of supercritical 
bifurcation for steadily moving spots. We consider the standard FitzHugh--Nagumo 
model including two variables -- a short-range activator $u$ and a long-range 
inhibitor $v$:
  \begin{eqnarray}
\epsilon^2 \tau u_t & = & \epsilon^2 \nabla^2 u + V'(u) - \epsilon v, 
      \label{sueq}   \\
 v_t & = & \nabla^2 v - v - \nu  + \mu u.
    \label{sveq}   \end{eqnarray}
Here $V(u)$ is a symmetric double-well potential with minima at $u=\pm 1$; 
$\epsilon \ll 1$ is a scale ratio, and other parameters are scaled in such a way 
that the effects of bias and  curvature on the motion of the front separating 
the up- and down states of the short-range variable are of the same order of 
magnitude. The local {\it normal} velocity of the front is 
\begin{equation}
c_n = \tau^{-1}(b v -\kappa) + O(\epsilon), 
\label{eqmot} \end{equation}
where $\kappa$ is curvature and 
$b$ is a numerical factor dependent on the form of $V(u)$; for example, 
$b=3/\sqrt{2}$ for the quartic potential $V(u)= -\frac{1}{4}(1-u^2)^2$. By 
definition, the velocity is positive when the down-state $u<0$ advances.

In the sharp boundary approximation valid at $\epsilon \ll 1$, a closed equation 
of motion for a solitary spot propagating with a constant speed can be written by 
expressing the local curvature in Eq.~(\ref{eqmot}) with the help of a suitable 
parametrization of the spot boundary, and resolving Eq.~(\ref{sveq}) rewritten 
in a coordinate frame propagating with a speed $c$ (as yet unknown). It is 
convenient to shift the long-range variable $v= w- \nu +\mu$, so that 
$w(\infty)=0$ when the up-state $u=1-O(\epsilon)$ prevails at infinity. The 
stationary equation of $w$ in the coordinate frame translating with the speed 
{\bf c} is 
\begin{equation}
  {\bf c} \cdot \nabla w + \nabla^2 w - w = 2\mu H,
\label{sweq}  \end{equation}
where, neglecting $O(\epsilon)$ corrections, $H=1$ inside and $H=0$ outside the 
spot. The solution can be presented in the form of an integral over the spot 
area $\cal S$:
\begin{equation}
w({\bf x} ) =  -\frac{\mu}{\pi}  \int_{\cal S} {\cal G}({\bf x}-\mbox{\boldmath 
$\xi$}) d^2 \mbox{\boldmath $\xi$},
\label{wint}  \end{equation}
where the kernel ${\cal G}$ contains a modified Bessel function $K_0$:
\begin{equation}
{\cal G}({\bf r})  = \frac{1}{2\pi} e^{-\frac{1}{2}{\bf c \cdot r}} 
  K_0\left( |{\bf r}|\sqrt{1+\mbox{$\frac{1}{4}c^2$}}\right) .
\label{wker}  \end{equation}
This integral can be transformed into a contour integral with the help of the 
Gauss theorem. To avoid divergent expressions, the contour should exclude the 
point ${\bf x}=\mbox{\boldmath $\xi$}$. Clearly, excluding an infinitesimal 
circle around this point does not affect the integral (\ref{wint}), since the 
kernel (\ref{wker}) is only logarithmically divergent. Replacing ${\cal G}({\bf 
r}) =\nabla^2 {\cal G}({\bf r}) + {\bf c} \cdot \nabla {\cal G}({\bf r})$ $({\bf r} 
\neq 0)$, we transform the integral in Eq.~(\ref{wint}) as
\begin{eqnarray}
-\int_{\cal S} {\cal G}({\bf x}-\mbox{\boldmath $\xi$}) 
d^2 \mbox{\boldmath $\xi$} =
\int_{\cal S} \nabla_{\mbox{\boldmath $\xi$}} \cdot 
{\bf H} ({\bf x}-\mbox{\boldmath $\xi$}) d^2 \mbox{\boldmath $\xi$} \cr
 = \oint_{\Gamma'} {\bf n}(s) \cdot {\bf H}
({\bf x}-\mbox{\boldmath $\xi$}(s)) ds, 
\label{gauss}  \end{eqnarray}
where ${\bf H}({\bf r}) = \nabla{\cal G}({\bf r}) + {\bf c} {\cal G}({\bf r})$ 
and {\bf n} is the normal to the contour $\Gamma'$. The vector Green's function  
{\bf H} corresponding to the kernel in Eq.~(\ref{wint}) is computed as
\begin{eqnarray}
 {\bf H}({\bf r}) &=& e^{-\frac{1}{2}{\bf c}\cdot {\bf r}} 
\left[\mbox{$\frac{1}{2}$}{\bf c}
K_0\left( |{\bf r}| \sqrt{1+\mbox{$\frac{1}{4}$} c^2}\right) \right. \cr
&-& \left. \sqrt{1+\mbox{$\frac{1}{4}$} c^2} \frac{{\bf r}}{|{\bf r}|}
K_1\left( |{\bf r}| \sqrt{1+\mbox{$\frac{1}{4}$} c^2}\right) \right].
\label{gauss2}  \end{eqnarray}
When ${\bf x}$ is a boundary point, $\Gamma'$ consists of the spot boundary 
$\Gamma$ cut at this point and closed by an infinitesimally small semicircle 
about ${\bf x}$. The integral over the semicircle equals to $\pi$. Defining the 
external normal to $\Gamma$ as the tangent ${\bf t}={\bf x}'(s)$ rotated 
clockwise by $\pi/2$, the required value of the long-range variable on the spot 
boundary (parametrized by the arc length $s$ or $\sigma$) is expressed,
using the 2D cross product $\times$, as
\begin{equation}
 v (s) = -\nu + \frac{\mu}{\pi}  \oint_{\Gamma} 
   {\bf H} ({\bf x}(s)- {\bf x} (\sigma)) \times {\bf x}' (\sigma) d\sigma .
\label{woint}  \end{equation}

To obtain a closed integral equation of a steadily moving spot, it remains to 
define a shift of parametrization accompanying shape-preserving translation. 
Recall that Eq.~(\ref{eqmot}) determines the propagation velocity $c_n$ along 
the {\it normal} to the boundary. In addition, one can introduce arbitrary {\it 
tangential} velocity $c_t$ which has no physical meaning but might be necessary 
to account for the fact that each ``material point'' on a translated contour is, 
generally, mapped onto a point with a different parametrization even when the 
shape remains unchanged. The tangential velocity can be defined by requiring 
that each material point be translated strictly parallel to the direction of 
motion, i.e. $c_n{\bf n}+c_t{\bf t}={\bf c}$. Taking the cross product with 
${\bf c}$ yields $c_t= c_n ({\bf c} \times {\bf t})$ $/ ({\bf c} \cdot {\bf 
t}).$ Then eliminating $c_t$ gives the normal velocity $c_n = {\bf c} \times 
{\bf t}$ necessary for translating the contour along the $x$ axis with the 
velocity $c$. Using this in Eq.~(\ref{eqmot}) yields the condition of stationary 
propagation
\begin{equation}
{\bf c} \times {\bf x}'(s) = \tau^{-1} [b v(s)- \kappa(s)] .
  \label{shape}  \end{equation}

The form and the propagation speed of a slowly moving and weakly distorted 
circular contour can be obtained by expanding Eq.~(\ref{shape}) in $c=|{\bf c}|$ 
near the point of traveling bifurcation $\tau=\tau_0$, which is also determined 
in the course of the expansion. For a circular contour with a radius $a$, 
Eq.~(\ref{woint}) takes the form 
\begin{eqnarray}
 v(\phi) &=& -\nu + \frac{\mu a}{\pi}  \int_{0}^{2\pi} 
   e^{-\frac{1}{2} c a(\cos \phi- \cos \varphi)} \; \times \cr
&& \left[ \mbox{$\frac{1}{2}$} c \cos \varphi \,
 K_0\left((2a\sqrt{1+\mbox{$\frac{1}{4}c^2$}}
\sin \mbox{$\frac{1}{2}$} |\phi-\varphi| \right) \right. \cr
& & + \sin \mbox{$\frac{1}{2}$}|\phi-\varphi|
\sqrt{1+\mbox{$\frac{1}{4}c^2$}}\, \; \times \cr 
&& \left. K_1\left(2a\sqrt{1+\mbox{$\frac{1}{4}c^2$}}
\sin \mbox{$\frac{1}{2}$} |\phi-\varphi| \right)  \right] d \varphi,
\label{woint1}  \end{eqnarray}
where $\phi$ or $\varphi$ is the polar angle counted from the direction of 
motion. The angular integrals that appear in the successive terms of the 
expansion are evaluated iteratively, starting from $\Phi_0(a)= \pi I_0(a) 
K_0(a)$ and using the relations
\begin{eqnarray*}
\Psi_k(a) &=& \int_{0}^{\pi} \sin^{2k+1} \frac{\phi}{2}\,
  K_1\left( 2 a\sin \frac{\phi}{2}\right) d \phi = 
- \frac{1}{2}\frac{d \Phi_k}{da}, \cr 
\Phi_k (a) &=& \int_{0}^{\pi}\!\! \sin^{2k} \frac{\phi}{2}\,
  K_0 \!\left( 2 a\sin \frac{\phi}{2} \right) \! d\phi = 
-\frac{1}{2a}\, \frac{d(a\Psi_{k-1})}{da}. 
\end{eqnarray*}

Effect of small boundary distortions on $v$ can be computed directly with the 
help of Eq.~(\ref{wint}), where the integration should be carried out only over 
a small area swept by the displaced spot boundary. This approach is most useful 
for stability analysis with respect to small perturbations of a known static 
shape, and is easier than using the expansion of Eq.~(\ref{woint}) with a 
perturbed boundary. For a circular spot, we expand the perturbations of both $v$ 
and $\rho$ in the Fourier series 
\begin{eqnarray}
\widetilde \rho(\phi,t) &=& \rho(\phi)-a =
 \sum_{n \geq 2} c^n a_n e^{\lambda_n t}\cos n\phi , \cr
 \widetilde v(\phi,t) &=& 
\sum_{n \geq 2} \widehat v_n e^{\lambda_n t}\cos n\phi .
\label{expand}  \end{eqnarray}
 The curvature is expressed as
\begin{eqnarray}
\kappa(\phi) &=& \frac{\rho^2 -2 \rho^2_\phi - 
          \rho\rho_{\phi\phi}}{(\rho^2+\rho^2_\phi)^{3/2}} \cr
&=& a^{-1}+ 3(c/a)^2 a_2 e^{\lambda_2 t}\cos 2\phi + O(c^3). 
\label{curv}  \end{eqnarray}

\begin{figure}[b]
\centerline{\hspace{.0cm} \psfig{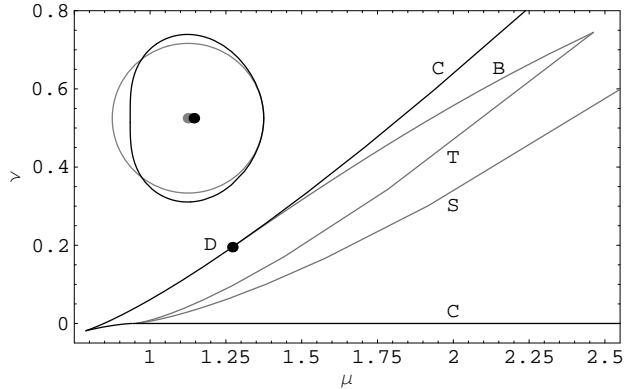}}
\caption{
The bifurcation diagram for stationary spots at $\tau=1$. C -- existence 
boundary, S -- locus of splitting instability. The stability region is bounded by 
the locus of breathing instability B, branching off at the point of double zero 
eigenvalue D, and the locus of traveling instability T. Inset: a circular spot 
distorted by second and third harmonics with amplitudes proportional to $c^n$. 
The shape is characteristic to a spot propagating to the right, and the 
amplitudes are chosen in such a way that the curvature on the back side 
vanishes. The center of the gray circle 
is shifted from the black to the gray spot to compensate 
the distortion at $\phi=0$, so that the integral   
is taken over the area between the black contour and the gray circle when the 
effect of small distortions on the $v$ field at this point is computed. 
\label{f1}} 
\end{figure}

Since the displaced point should remain on the boundary, the distortion 
$\widetilde \rho(\varphi)$ should be compensated by rigid displacement of the 
spot by an increment $\widetilde \rho(\phi)$ when $\widetilde v(\phi)$ is 
computed (see the inset in Fig.~\ref{f1}). The resulting equation for 
eigenvalues $\lambda_n$ following from Eq.~(\ref{eqmot}) is
\begin{eqnarray}
&& \tau \lambda_n = \frac{ n^2-1}{a^2} - \frac{4ab\mu}{\pi^2} 
\int_0^{\pi} \cos n\phi \, d \phi \; \times \cr
&& \int_0^\pi  [\widetilde \rho(\varphi)- \widetilde \rho(\phi)
\cos (\varphi-\phi)] 
e^{-\frac{1}{2} c a(\cos \phi- \cos \varphi)}\; \times \cr
&& K_0\left(2a\sqrt{1+ \lambda_n +\mbox{$\frac{1}{4}c^2$}}
\sin \mbox{$\frac{1}{2}$} |\phi-\varphi| \right) d \varphi.
\label{wint1}  \end{eqnarray}

Using the constant zero-order term in the expansion of Eq.~(\ref{woint1}) 
together with $\kappa= a^{-1}$ in Eq.(\ref{eqmot}) yields the stationarity 
condition 
\begin{equation}
 \nu = -(ba)^{-1} + \mu a \left[ K_1(a)\,I_0(a) - K_0(a)\,I_1(a) \right] .
\label{cstat} \end{equation}
A stationary solution stable against collapse or uniform swelling exists in the 
region in the parametric plane $\mu,\nu$ (Fig.~\ref{f1}) bounded by the cusped 
curve C and the axis $\nu=0,\, \mu>2/b$. This curve is drawn as a parametric 
plot with $\nu(a)$ given by Eq.~(\ref{cstat}) and $\mu(a)$ by Eq.~(\ref{wint1}) 
with $c, \, n$ and $\lambda_0$ set to zero (or, equivalently, by the condition 
$F_0'(a)=0$, where $F_0(a)$ is the right-hand side of Eq.~(\ref{cstat}).

The first-order term in the expansion of Eq.~(\ref{woint1}) is proportional to 
$\cos \phi$, and should compensate at the traveling bifurcation point the 
left-hand side of Eq.~(\ref{shape}). This yields the bifurcation condition 
\begin{equation}
\tau_0  =  b\mu a [a(I_1(a)K_0(a) - I_0(a)K_1(a)) + 2I_1(a)K_1(a)] ,
  \label{disp1d} \end{equation}
which coincides with the known result obtained by other means \cite{mikh}.
The curve T in Fig.~\ref{f1} shows the traveling instability threshold for 
$\tau_0  =1$. The static spot is unstable below this curve; the locus shifts up 
(to smaller radii) as $\tau$ decreases, and exits the existence domain 
at $\tau<1/4$. At $\tau>1$, the dominant instability at large radii is a static 
splitting instability. Its locus, determined by Eq.~(\ref{wint1}) with $n=2$ and 
$c=\lambda_2=0$, is the curve S in Fig.~\ref{f1}.
 
Another possible dynamic instability is breathing instability 
\cite{ohta,haim,pur99}. Its locus is given by Eq.~(\ref{wint1}) with $c=n=0$ and 
$\lambda_0=i\omega$. The frequency $\omega$ as a function of the spot radius $a$ 
is computed by solving the equation $\tau \omega =a^{-2}$Im $F(a,\omega)/$Re  
$F(a,\omega)$, where $F(a,\omega)$ is the right-hand side of Eq.~(\ref{wint1}) 
computed as
\begin{eqnarray}
F(a,\omega)  &=& 2 \mu a \left[I_1\left(a\sqrt{1+i\omega}\right)
K_1\left(a\sqrt{1+i\omega}\right) \right. \cr
&-& \left. I_0\left(a\sqrt{1+i\omega}\right)
K_0\left(a\sqrt{1+i\omega}\right) \right] . 
  \label{dis00} \end{eqnarray}
The curve B in Fig.~\ref{f1} shows the bifurcation locus at $\tau=1$. The 
instability region retreats to small radii (large $\nu$) at large $\tau$ and 
spreads downwards as $\tau$ decreases. The balloon of stable solutions disappears 
altogether at $\tau<0.5$ after the tips of both dynamic loci meet on the 
existence boundary.

In the second order, Eq.~(\ref{woint1}) yields a constant term 
\begin{equation}
v^{(2,0)}  = - \mu a^2 [a(I_1(a)K_0(a) - I_0(a)K_1(a)) + I_1(a)K_1(a)]  
  \label{dis20} \end{equation}
and a dipole term $ v^{(2,2)} = q^{(2,2)} \cos 2\phi$, where
\begin{eqnarray}
q^{(2,2)} = \mbox{$\frac{1}{4}$} \mu a^2 
 [a(I_0(a)K_1(a) - I_1(a)K_0(a)) \cr
- 3I_1(a)K_1(a) +2 I_2(a)K_2(a)] .
  \label{dis22} \end{eqnarray}
The constant term is positive and causes contraction of the average radius of 
the moving spot by an increment $\widetilde a = - a^2 c^2 bv^{(2,0)}$. 

The second-order dipolar term in the right-hand side of Eq.~(\ref{shape}), 
$\widetilde v^{(2,2)} = \widetilde q^{(2,2)} a_2 \cos 2\phi$, as well as the 
third-order first harmonic term, $\widetilde v ^{(3,1)} = \widetilde q^{(3,1)} 
a_2 \cos \phi$, needed for the solvability condition to follow, are read from 
Eq.~(\ref{wint1})  with $n=2$ and $\lambda_2=0$, respectively, in zero and first 
order in $c$:
\begin{eqnarray}
\widetilde q^{(2,2)} &=&  - 3a^{-2} + 2b\mu [I_1(a)K_1(a)- I_2(a)K_2(a)],
  \label{disp22} \\
\widetilde q^{(3,1)} &=& b\mu a^2 I_1(a)K_1(a) .
  \label{disp31} \end{eqnarray}
The coefficient $\widetilde q^{(2,2)}$ vanishes at the splitting instability 
threshold (curve S in Fig.~\ref{f1}), and must be negative when the circular 
spot is stable. Consequently, the distortion amplitude is $a_2 = - q^{(2,2)}/ 
\widetilde q^{(2,2)} <0$, so that the dipole term causes contraction of the 
moving spot in the direction of motion and expansion in the normal direction. 

Continuing the expansion to the third order, we compute the first harmonic term 
contributing to the solvability condition. The latter has the form $ \widetilde 
\tau c = k c^3$, where $ \widetilde \tau = \tau-\tau_0$ and the coefficient $k$
determining the character of the bifurcation is computed as
\begin{equation}
k = b\mu \left( q^{(3,1)} - \tau_0'(a) a^2 v^{(2,0)} 
- \widetilde q ^{(3,1)} q ^{(2,2)}/ \widetilde q ^{(2,2)}\right).
  \label{disp3} \end{equation}
The first term is the coefficient at the first harmonic in the third order of 
the expansion of Eq.~(\ref{woint1}). The second term takes into account the 
second-order radius correction to the first-order first harmonic term. The last 
term gives the effect of dipolar shape distortion; it becomes dominant when the 
locus of splitting instability is approached. Stable traveling solution should 
be observed beyond the traveling instability threshold, i.e.\ at $\widetilde 
\tau <0$; hence, the condition of supercritical bifurcation is $k<0$. The 
numerical check of the symbolically computed expression shows that the traveling 
bifurcation is always supercritical when the spot is stable to splitting. The
traveling solution bifurcating supercritically must be stable, at least close to the
bifurcation point where it inherits stability of the stationary spot to other kinds
of perturbations.

The third harmonic term that appears in the third order of the expansion 
delineates, together with the second-order dipolar term, the characteristic 
shape of a translating spot, pointed in the direction of motion and spread 
sidewise, as in the inset in Fig.~\ref{f1}, 
which has been also observed in numerical 
simulations \cite{bode}. Beyond the range of the bifurcation expansion, the 
shape, as well as the propagation speed can be determined by solving numerically 
Eq.~(\ref{shape}) with $v(s)$ given by Eq.~(\ref{gauss2}) and curvature computed
using  the 
fully nonlinear expression in Eq.~(\ref{curv}). Although the boundary integral 
method reduces a PDE to a 1D integro-differential equation, the equation is 
rather difficult. Iterative numerical solution \cite{dima} tends to break down 
rather close to the bifurcation point, as soon as the shape distortion becomes 
strong enough to flatten the spot at the back side. Since the boundary integral 
equation is non-evolutionary, there is no way to distinguish between a purely 
numerical failure of convergence and a physical instability  that would lead to 
lateral spreading observed in PDE simulations \cite{mikh}. 

The above bifurcation expansion proves that a stable traveling solution does 
exist in the basic model (\ref{sueq}), (\ref{sveq}) in the sharp boundary limit. 
The result is applicable at $1 \gg c\gg \sqrt{\epsilon}$. It can be extended 
straightforwardly to models with more than one long-range variable, provided all 
long-range equations are linear. Stable traveling spot solutions should be, 
indeed, more robust in an extended model where they have been obtained in PDE 
simulations \cite{bode}, whereas in the basic model they require fine parametric 
tuning aided by the analytical theory.

\paragraph*{Acknowledgement.} 
This work has been supported by the German--Israeli Science
Foundation.

\end{document}